\documentclass[twocolumn]{aastex63}

\received{}
\revised{}
\accepted{}
\submitjournal{ApJL}

\shorttitle{Anisotropy of Inner-Heliospheric Kinetic Turbulence}
\shortauthors{Duan et al.}
\graphicspath{{./}{figures/}}

\begin{document}

\title{Anisotropy of Solar-Wind Turbulence in the Inner Heliosphere at Kinetic Scales: PSP Observations}

\correspondingauthor{Jiansen He}
\email{jshept@pku.edu.cn}
\correspondingauthor{Stuart D. Bale}
\email{bale@berkeley.edu}

\author[0000-0002-6300-6800]{Die Duan}
\affil{School of Earth and Space Sciences, Peking University, Beijing, 100871, China}
\author[0000-0001-8179-417X]{Jiansen He}
\affil{School of Earth and Space Sciences, Peking University, Beijing, 100871, China}
\author[0000-0002-4625-3332]{Trevor A. Bowen}
\affil{Space Sciences Laboratory, University of California, Berkeley, CA 94720-7450, USA}
\author[0000-0003-2845-4250]{Lloyd D. Woodham}
\affiliation{The Blackett Laboratory, Imperial College London, London, SW7 2AZ, UK}
\author[0000-0003-3072-6139]{Tieyan Wang}
\affiliation{RAL Space, Rutherford Appleton Laboratory, Harwell Oxford, Didcot OX11 0QX, UK}
\author{Christopher H. K. Chen}
\affil{School of Physics and Astronomy, Queen Mary University of London, London E1 4NS, UK}
\author[0000-0001-9202-1340]{Alfred Mallet}
\affil{Space Sciences Laboratory, University of California, Berkeley, CA 94720-7450, USA}
\author[0000-0002-1989-3596]{Stuart D. Bale}
\affiliation{Space Sciences Laboratory, University of California, Berkeley, CA 94720-7450, USA}
\affiliation{Physics Department, University of California, Berkeley, CA 94720-7300, USA}
\affiliation{The Blackett Laboratory, Imperial College London, London, SW7 2AZ, UK}
\affiliation{School of Physics and Astronomy, Queen Mary University of London, London E1 4NS, UK}

\begin{abstract}

The anisotropy of solar wind turbulence is a critical issue in understanding the physics of energy transfer between scales and energy conversion between fields and particles in the heliosphere. Using the measurement of \emph{Parker Solar Probe} (\emph{PSP}), we present an observation of the anisotropy at kinetic scales in the slow, Alfv\'enic, solar wind in the inner heliosphere. \textbf{The magnetic compressibility behaves as expected for kinetic Alfv\'enic turbulence below the ion scale.} A steepened transition range is found between the inertial and kinetic ranges in all directions with respect to the local background magnetic field direction. The anisotropy of $k_\perp \gg k_\parallel$ is found evident in both transition and kinetic ranges, with the power anisotropy $P_\perp/P_\parallel > 10$ in the kinetic range leading over that in the transition range and being stronger than that at 1 au. The spectral index varies from $\alpha_{t\parallel}=-5.7\pm 1.0$ to $\alpha_{t\perp}=-3.7\pm 0.3$ in the transition range and $\alpha_{k\parallel}=-3.12\pm 0.22$ to $\alpha_{k\perp}=-2.57\pm 0.09$ in the kinetic range. The corresponding wavevector anisotropy has the scaling of $k_\parallel \sim k_\perp^{0.71\pm 0.17}$ in the transition range, and changes to $k_\parallel \sim k_\perp^{0.38\pm 0.09}$ in the kinetic range, consistent with the kinetic Alfv\'enic turbulence at sub-ion scales.

\end{abstract}


\section{Introduction} \label{sec:intro}
Magnetic field fluctuations in the solar wind are highly turbulent. The measured power spectral density (PSD) of the fluctuating magnetic field always exhibits power laws $k^{-\alpha}$, where $k$ is the wavenumber, and $\alpha$ is the spectral index. A single spacecraft measures the PSD as a function of $f^{-\alpha}$ in the frequency domain, which can be converted to the spatial domain under the Taylor Hypothesis. According to the physical processes at different scales, the PSD in the solar wind can be divided into several segments, which can be fitted with different $\alpha$. The inertial range, which is dominated by magnetohydrodynamic (MHD) turbulence, follows the cascade models with spectral indices $\alpha_i$ from around 3/2 to 5/3 \citep{Bruno2013, Chen2020APJS}. The PSDs become steepened below the ion scales (ion thermal gyroradius $\rho_i$ or ion inertial length $d_i$), where kinetic mechanisms should be taken into account. Sometimes a sharp transition range is observed with $\alpha_t\sim 4$ \citep{Sahraoui2010, Bowen2020inner}. \textbf{This transition range may be caused by imbalanced turbulence \citep{Voitenko2016, meyrand2020violation}, energy dissipation of kinetic waves \citep{Howes2008}, ion-scale coherent structures \citep{Lion2016}, or a reconnection dominated range \citep{Mallet2017}. At smaller scales, a flatter sub-ion kinetic range forms with the spectral index $\alpha_k \sim 7/3$, which can be explained as the MHD Alfv\'enic turbulence developing into a type of kinetic wave turbulence, e.g., kinetic Alfv\'en waves (KAWs) \citep{Schekochihin2009} or whistler waves \citep{Cho2004}. Intermittency in the kinetic range could lead to an -8/3 spectrum \citep{Boldyrev2012, zhao2016kinetic}.  Ion-cyclotron-wave (ICW) turbulence could lead to a steeper -11/3 spectrum \citep{Krishan2004,Galtier2007,Meyrand2012,Schekochihin2019}. The kinetic range always behaves as the KAW turbulence with the slope of -2.8 in the near-earth space \citep{Bale2005a, Chen2013, Chen2016}. The spectral indices increase again beyond the electron kinetic scales in observations, indicating the conversion of turbulence energy to electrons \citep{Sahraoui2009, Alexandrova2012, Chen2019NC} or transitions to a further cascade \citep{Schekochihin2009,Chen2017}. In simulations, \citet{meyrand2013anomalous} obtained a -8/3 spectrum at electron scales under the 3D electron-MHD model.}

Because of the background interplanetary magnetic field (IMF), the turbulence in the solar wind is anisotropic. At the MHD scales, the energy transfer rate depends on the angle $\theta_{kB}$ between the wavevector $\mathbf{k}$ of fluctuations and the background magnetic field \citep{Goldreich1995}. The anisotropic energy cascade leads to the anisotropy of power level and spectral index \citep{Chen2010interpret}, which is observed in the solar wind turbulence \citep{Horbury2008,Podesta2009}. \citet{Goldreich1995} also predicts a critical balanced wavevector anisotropy of $k_\parallel\sim k_\perp^{2/3}$ and \citet{Boldyrev2006} predicts $k_\parallel\sim k_\perp^{1/2}$. Here $k_\perp$ is the wavevector perpendicular to the background magnetic field direction, and $k_\parallel$ is the wavevector along the parallel direction.  \citet{He2013} found that turbulent power is enhanced along a ridge at $k_\perp > k_\parallel$ in the 2D wavevector space. Moreover, it is argued that other possible reasons could lead to the observed anisotropy, such as intermittency \citep{Wang2014}, solar wind expansion \citep{Verdini2019} and non-stationarity of the background magnetic field \citep{Wu2020}. How the MHD-scale anisotropy rises in the solar wind is still a challenging question.

In the kinetic range, the fluctuations remain anisotropic. Theoretically, the specific form of wavevector anisotropy will depend on the nature of the fluctuations. The kinetic Alfv\'enic turbulence models predict $k_\parallel \sim k_\perp^{1/3}$ \citep{Howes2008, Schekochihin2009}. The intermittent KAW model gives the scaling of $k_\parallel \sim k_\perp^{2/3}$ \citep{Boldyrev2012}. The tearing-instability-mediated-turbulence model predicts from $k_\parallel \sim k_\perp^{2/3}$ to $k_\parallel\sim k_\perp$ \citep{Boldyrev2019}. In observations, the power along quasi-perpendicular directions are found dominant via the structure function approach \citep{Chen2010} and the $k$-filtering technique \citep{Sahraoui2010}. The wave modes are also anisotropic, as \citet{He2011} and \citet{Huang2020z} found that the ion-scale turbulence contains quasi-parallel Alfv\'en-cyclotron waves (ACWs) and quasi-perpendicular KAWs. The numerical kinetic simulation is another way to explore the physics of anisotropy, and different scalings are reached, for example, $k_\parallel\sim k_\perp$ \citep{Arzamasskiy2019, Landi2019}, $k_\parallel\sim k_\perp^{1/3}$ \citep{Groselj2018, Groselj2019} and $k_\parallel\sim k_\perp^{2/3}$ \citep{Cerri2019}.

The previous studies are mainly based on measurements in the vicinity of 1 au. \textbf{In the inner heliosphere, the \emph{Parker Solar Probe (PSP)} spacecraft \citep{Fox2016} encountered slow and high cross-helicity solar wind with low plasma $\beta$ at its first perihelion near 0.17 au \citep{Bale2019}. The fluctuating magnetic spectra always contain a strongly steepened transition range with $\alpha_t\sim 4$ below the ion scales \citep{Bowen2020inner}, which is seldom observed at 1 au. Investigating the anisotropy in and below the transition range here can provide us more information about the kinetic turbulence in this different parameter regime.} The paper is organized as: Section \ref{sec:data} describes the data and method used in this work. Section \ref{sec:result} shows the result of the anisotropy. Section \ref{sec:conclusion} is the conclusion and discussion.

\section{Data and Method} \label{sec:data}
The data from \emph{PSP} at its first perihelion (0.17 au) are used in this study. The FIELDS and Solar Wind Electron Alpha and Proton (SWEAP) instruments provide \emph{in situ} measurements of the inner-heliospheric solar wind \citep{Bale2016, kasper2016solar}. We use a merged data set from flux-gate magnetometer (FIELDS/MAG) and search coil (FIELDS/SCM) measurements (both operate at 293 Hz), resolving the full range from MHD to kinetic scales simultaneously \citep{bowen2020merged}. The plasma measurements are from the Solar Probe Cup (SWEAP/SPC) \citep{Case2020}. \textbf{We use the solar wind velocity in the spacecraft frame as the sampling direction.} During the perihelion, \emph{PSP} encountered a slow ($V_{SW}$ $<$ 400 km/s), but highly Alfv\'enic solar wind ($\sigma_c \sim 0.7$). The background radial magnetic field is anti-sunward \citep{Bale2019}.

The Morlet wavelet transform is employed to build the PSD of the magnetic fluctuations \citep{Horbury2008,Podesta2009}, located at 139 logarithmically spaced frequencies from 0.01 Hz to 149.5 Hz in the spacecraft frame. Part of the inertial range is defined at $0.1\ \mathrm{Hz} < f < 1\ \mathrm{Hz}$, as the ion-scale break frequency is usually larger than 1 Hz at 0.17 au \citep{Duan2020}. The power of the reaction wheels set on the spacecraft contaminates the power spectra around $20\sim 30$ Hz, so the kinetic range is defined as $40\ \mathrm{Hz} < f < 90\ \mathrm{Hz}$. A short-time-Fourier-transfrom method is used to remove the artificial spikes \citep{Bowen2020icw} (Woodham et al., in preparation). We avoid fitting $f>90\ \mathrm{Hz}$ ranges to avoid SCM noise floor \citep{bowen2020merged}. A piecewise linear fitting in log-log space is implemented to locate the transition range, which is described in Appendix \ref{App}. 

Gaussian windows are used to evaluate the local mean magnetic field directions at different scales, and the angles between the local magnetic field direction and average solar wind velocity direction $\theta_{BV}(f,t)$ are calculated. To estimate the angular distribution of PSD, we partition $\theta_{BV}(f,t)$ into 18 angle bins from $\theta_{BV} \in (90^\circ,95^\circ]$ to $\theta_{BV} \in (175^\circ,180^\circ]$.\textbf{ However, ICWs are common in the inner heliosphere, contributing to the formation of a bump around the ion scale in the power spectra when the solar wind velocity is (quasi-)(anti-)parallel to the magnetic field \citep{Bowen2020icw}. We identify the ICWs according to the reduced magnetic helicity along the radial direction with $\theta_{BV}>120^\circ$ and $\sigma_{mTN} > 0.5$ \citep{He2011}. The PSD is averaged over each bin as
\begin{equation}
    P(f,\theta_{i})=\frac{1}{N_{f,i}}\sum P(f,t)|_{\theta_i< \theta_{BV}(f,t)\leq\theta_i+5^\circ,\sigma_{mTN}(f,t)<0.5},
\end{equation}
where $\theta_i=5^\circ i+85^\circ, i=1,2,...,18$ and $N_{f,i}$ is the number of points without ICWs in the frequency and time domain \citep{Podesta2009}.}

\section{Results} \label{sec:result}
\begin{figure}[htb!]
    \centerline{\includegraphics{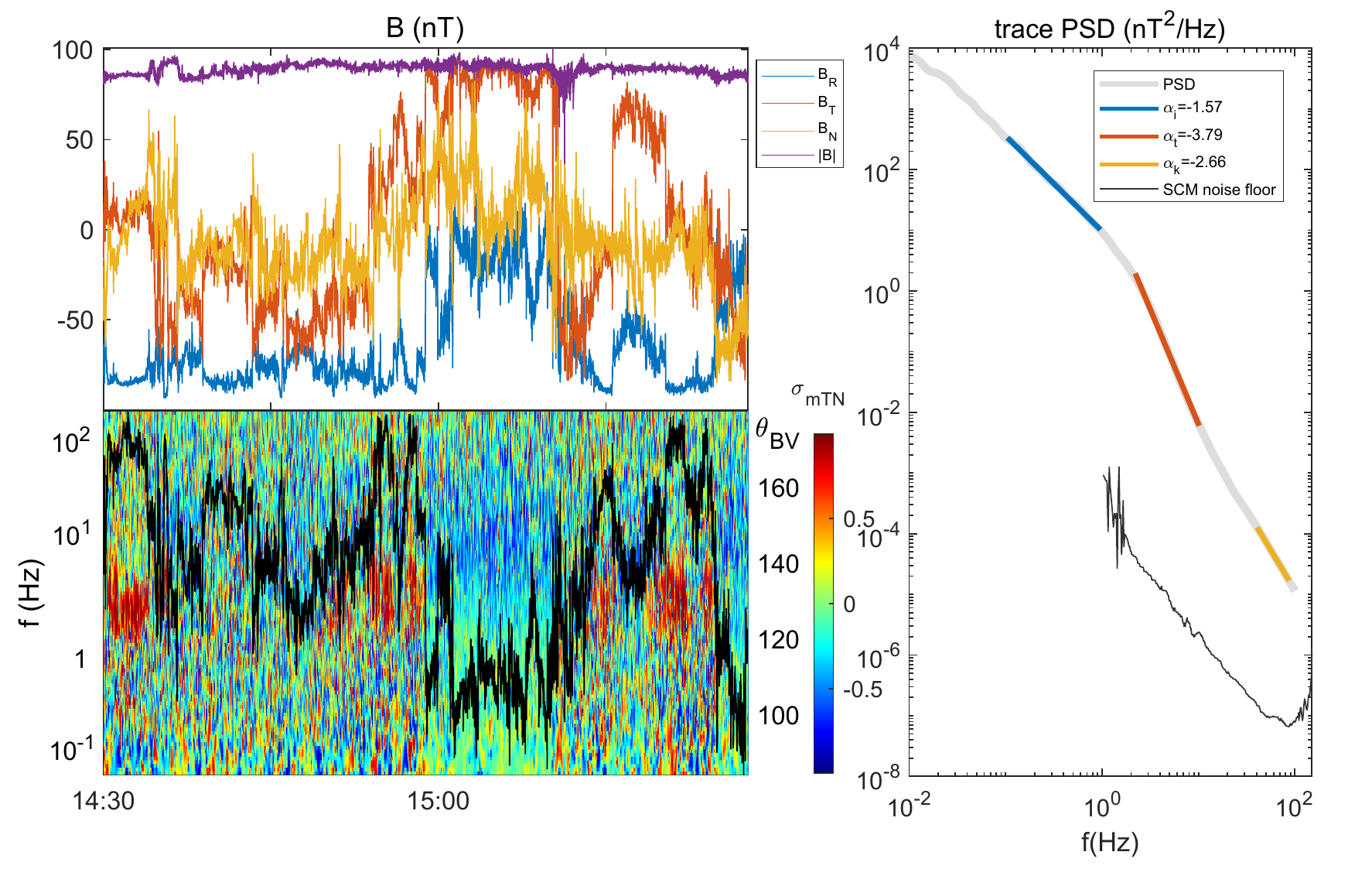}}
    \caption{The overview of the interval from 14:30 to 15:30 on Nov 6, 2018. (a) The magnetic field in the RTN coordinates. (b) The reduced magnetic helicity $\sigma_{mTN}$ along the radial direction. Positive values indicate LH modes and negative values indicate RH modes. The black line is angle between the magnetic field and the solar-wind-velocity direction in the spacecraft frame $\theta_{BV}$. (c) The averaged trace PSD (grey ) over the interval. The dash lines are the linear fittings to inertial (blue), transition (red) and kinetic (yellow) ranges. The bottom grey line is the noise level of the SCM.}
    \label{fig1}
\end{figure}

 Figure \ref{fig1} shows an example interval from 14:30 to 15:30 on Nov 5, 2018. The merged data set is in the Radial-Tangential-Normal (RTN) coordinate system, where $B_R$ is the radial component of the magnetic field along the Sun-spacecraft line. The amplitude of the magnetic field keeps constant as $|B|\sim 89\ \rm{nT}$. \textbf{The average proton density is $n_p\sim 316\ \rm{cm^{-3}}$, the solar wind speed is $V_{sw}\sim 342$ km/s, and the average proton thermal speed is $w_p\sim 61 $ km/s, yielding the Alfv\'en speed $v_A \sim 109$ km/s, the proton thermal gyroradius $\rho_p\sim 7$ km, the proton inertial length $d_p \sim 13$ km and the proton sound gyroradius $\rho_s\sim 6.2$ km. The plasma beta for protons and electrons are $\beta_p \sim 0.3$ and $\beta_e\sim 0.6$, and the average ratio of proton temperature to electrons is $T_p/T_e\sim 0.7$. The interval is highly imbalanced with $\sigma_c\sim 0.9$.} The corresponding frequencies of the electron scales is higher than the Nyquist frequency. The $\theta_{BV}$ covers the range from $90^\circ$ to $180^\circ$, allowing to estimate the anisotropy. The inertial, transition and kinetic ranges are observed distinctly in the averaged trace PSD. In the inertial range, the spectral index $\alpha_i$ is -1.56, similar to the statistical result of \citet{Chen2020APJS} at 0.17 au. Then the PSD sharply decreases with $\alpha_t=-3.77$ in the transition range. In the kinetic range, the spectral index increases to $\alpha_k=-2.67$, which is close to -8/3 but larger than -2.8 from studies near 1 au \citep{Sahraoui2009}. To explore the nature of the transition range, we calculate the normalized reduced magnetic helicity along the radial direction \citep{He2011, Woodham2018}. Positive helicity represents left-handed (LH) wave modes and negative helicity represent right-handed (RH) modes for sunward background magnetic field. It is revealed that there are two components with opposite polarization around 1 to 20 Hz. The LH modes, locating near 1 to 6 Hz when the magnetic field is quasi-parallel to the radial direction, are identified as coherent ion-scale cyclotron waves \citep{Bowen2020icw}. When $\theta_{BV}$ is close to $90^\circ$, the RH modes dominate around 4 to 20 Hz, which could be the quasi-perpendicular KAWs \citep{Huang2020z}. 

\begin{figure}[htb!]
    \centerline{\includegraphics{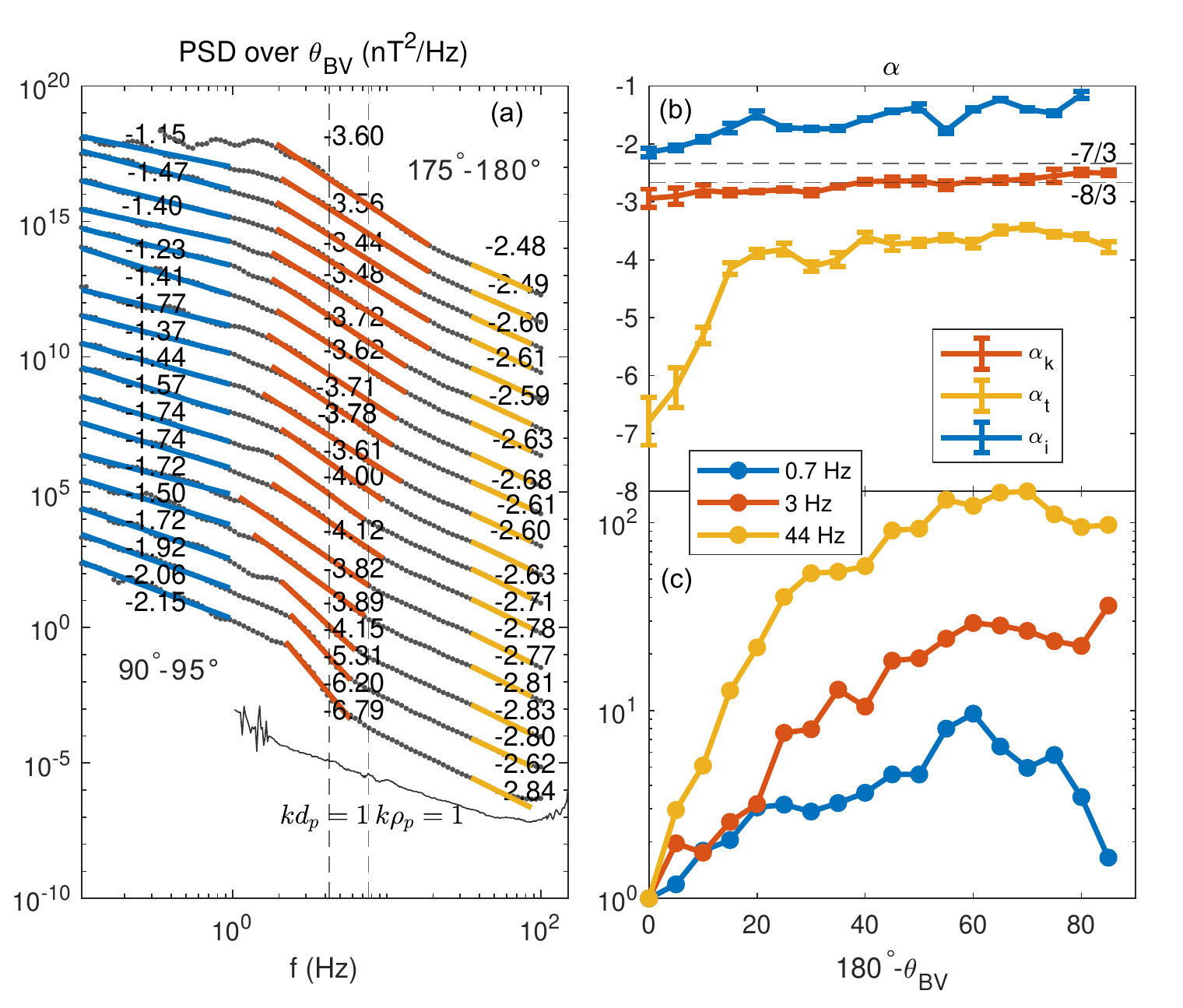}}
    \caption{(a) The magnetic PSDs in different angle bins. The colored dashed lines are the fitted inertial (blue), transition (red) and kinetic (yellow) ranges. Black spectrum is the noise level of the SCM. The spectral indices are also shown. The vertical dashed lines indicate the characteristic scales $d_i$ (yellow) and $\rho_i$ (purple). (b) Spectral indices of the three ranges estimated from different $\theta_{BV}$ bins. (c) The power anisotropy at three specific frequencies respectively located in the inertial (0.7 Hz, yellow), transition (3 Hz, red) and kinetic (44 Hz, blue) ranges over different $\theta_{i}$.}
    \label{fig2}
\end{figure}

The angular distribution of the PSDs $P(f,\theta_{i})$ are shown in Figure \ref{fig2}. From the bottom to top, the different curves correspond to different angular bins from the parallel to the perpendicular directions. The parallel spectrum is flattened with $f>60 \ \mathrm{Hz}$, where it reaches the noise level of the SCM. We only use the range of $40\ \mathrm{Hz}<f< 55\ \mathrm{Hz}$ to fit the parallel kinetic spectrum. The PSDs in other directions are larger than the noise level of the SCM, indicating the vadility of the measurement. The PSDs for the remaining angular bins have been offset by factors of 10 for easier viewing. We demonstrate for the first time that the transition range exists in all of the directions in the inner heliosphere. The break between the inertial range and the transition range $f_{it}$ is around 2 Hz, and the break between the transition and kinetic ranges $f_{tk}$ is near 5 to 20 Hz. Using Taylor's Hypothesis, we calculate the Doppler frequency corresponding to the scales of $\rho_i$ and $d_i$ in the spacecraft frame. We find that the frequencies of $f_{\rm{di}}=V_{\rm{sw}}/2\pi k$ with $kd_i\sim 1$ and $f_{\rm{\rho i}}=V_{\rm{sw}}/2\pi k$ with $k\rho_i\sim 1$ are sitting between the spectral break frequencies of $f_{it}$ and $f_{tk}$. Taylor hypothesis has been shown to hold in the inertial range for the early PSP orbits \citep{perez2021applicability}.

Figure \ref{fig2} (b) shows the spectral anisotropy for the three ranges. The spectral indices $\alpha$ of each range all have a decreasing trend from the quasi-parallel direction to the quasi-perpendicular direction. The spectral index $\alpha_i$ is -1.4 along the perpendicular direction and $\alpha_i\sim -2.1$ along the parallel direction, demonstrating a similar trend to that of critical-balanced anisotropy observed at 1 au \citep{Horbury2008, Podesta2009}. In the transition range, the PSD is steepened sharply with $\alpha_t\sim -6.8$ along the parallel direction, and changes to $\alpha_t\sim-3.6$ along the perpendicular direction. This spectral anisotropy has a similar angular dependence with the observation at 1 au (see Figure 4 in \citet{Duan2018}). However, at 1 au, the spectral index varies from around -4 to -2.8, much shallower than the inner heliosphere. Extending to the kinetic scales, the spectral index $\alpha_k$ increases to -2.8 at parallel direction along the parallel direction and -2.5 along the perpendicular direction, which is consistent with the anisotropy of the $\delta B_\perp$ spectra from the \textit{Cluster} observations \citep{Chen2010}. 

We define the perpendicular and parallel power spectra as $P_\perp (f)=P(f,90^\circ <\theta_{BV}\leq 95^\circ)$ and $P_\parallel (f)=P(f,175^\circ<\theta_{BV}\leq 180^\circ)$. Figure \ref{fig2} (c) shows the power spectra ratio ($P (f,\theta_{\rm{BV}})/P_\parallel (f)$, including $P_\perp (f)/P_\parallel (f)$) at three selected frequencies in the three ranges. The power anisotropy $P_\perp/P_\parallel$ is around 3 at 0.7 Hz in the inertial range, and increases to 30 at 3 Hz in the transition range. At the kinetic scales, $P_\perp/P_\parallel$ reaches 90 at 44 Hz, which is much larger than 5 measured by structure function in \citet{Chen2010}. It reveals that below the transition range, the power anisotropy at kinetic scales in the inner heliosphere is stronger than at 1 au.

\begin{figure}[htb!]
    \centerline{\includegraphics{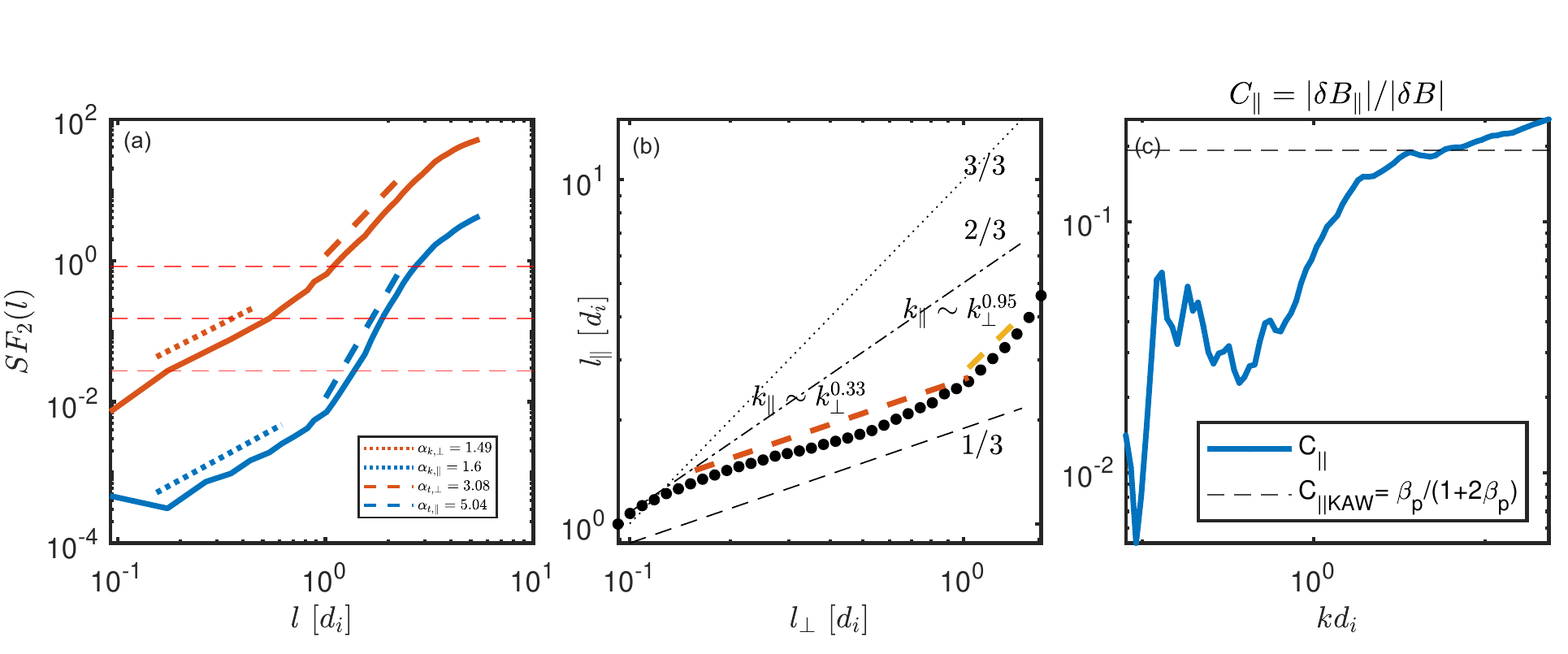}}
    \caption{(a) Structure functions along the parallel (blue) and perpendicular (red) directions. Dashed and dotted lines are the fitting results in both the transition and kinetic ranges. Horizontal dashed red lines indicate the range to calculate the wavevector anisotropy. (b) Wavevector anisotropy derived from (a). red and yellow lines are for the kinetic and transition ranges (along the perpendicular direction), respectively. Three typical relations are presented as black dot lines for reference. (c) The average magnetic compressibility in the interval. The black dashed line is the theoretical prediction for the KAW at sub-ion scales.}
    \label{fig3}
\end{figure}

Using the method from \citet{tieyanWang2020}, the five-points structure functions $SF_2(l)$ along the parallel and perpendicular directions are calculated and shown in Figure \ref{fig3} to explore the wavevector anisotropy, here $l = 1/k$ is the spatial displacement. The spectral indices $\alpha_{SF}$ of the structure functions are consistent with the indices $\alpha_{PSD}$ from the PSD as $|\alpha_{SF}|+1=|\alpha_{PSD}|$\citep{Chen2010}. 
By equating $SF_{2}(l_\parallel)$ and $SF_{2}(l_\perp)$, the anisotropy relation between $l_\parallel$ and $l_\perp$ is estimated. \textbf{Due to the strong power anisotropy, when the perpendicular structure function reaches the noise, the parallel structure function is still in the transition range. In the transition range (along the perpendicular direction) we get $l_\parallel \sim l_\perp^{0.95}$. Below $d_i$, $l_\parallel \sim l_\perp^{0.33}$ in the kinetic range, similar to the prediction of critical-balance KAW turbulence of $k_\parallel \sim k_\perp^{1/3}$ \citep{Schekochihin2009}. In principle, $l$ and $k$ have the same anisotropy scaling. We calculate the magnetic compressibility $C_\parallel = |\delta B_\parallel|/|\delta \mathbf{B}|$ to explore the nature of the sub-ion scales in Figure \ref{fig3}(c). The $C_\parallel < 0.1$ when $kd_i<1$, and increase to $0.2$ in the kinetic range, is similar to the $C_\parallel$ of KAWs, not whistler waves \citep{Salem2012}. In the kinetic range, $k_\parallel \sim k_\perp^{0.33}$ is close to the relation of the critical-balanced KAW turbulence with $k_\parallel \sim k_\perp^{1/3}$, but $\alpha_{k\perp}\approx -2.48 < -7/3$ is different from the KAW model, which is similar to the simulations of \citet{Groselj2018} and \citet{Groselj2019} with $\alpha_k\sim -2.8$ and $k_\parallel\sim k_\perp^{1/3}$.}

Statistical analysis of the wavevector anisotropy is performed by dividing the data during Nov 5 to Nov 7 into one-hour intervals with 50\% overlapping when the SCM was operating at 293 Hz. Here we only consider the transition and kinetic ranges, because several directions do not have enough samples in the inertial range. Only $\theta_{\rm{BV}}>90^\circ$ is considered. Intervals that do not have enough samples (counts $<$ 5000) in the perpendicular or parallel directions to provide spectra in both directions are also excluded. We get 22 intervals in total.

\begin{figure}[htb!]
    \centerline{\includegraphics{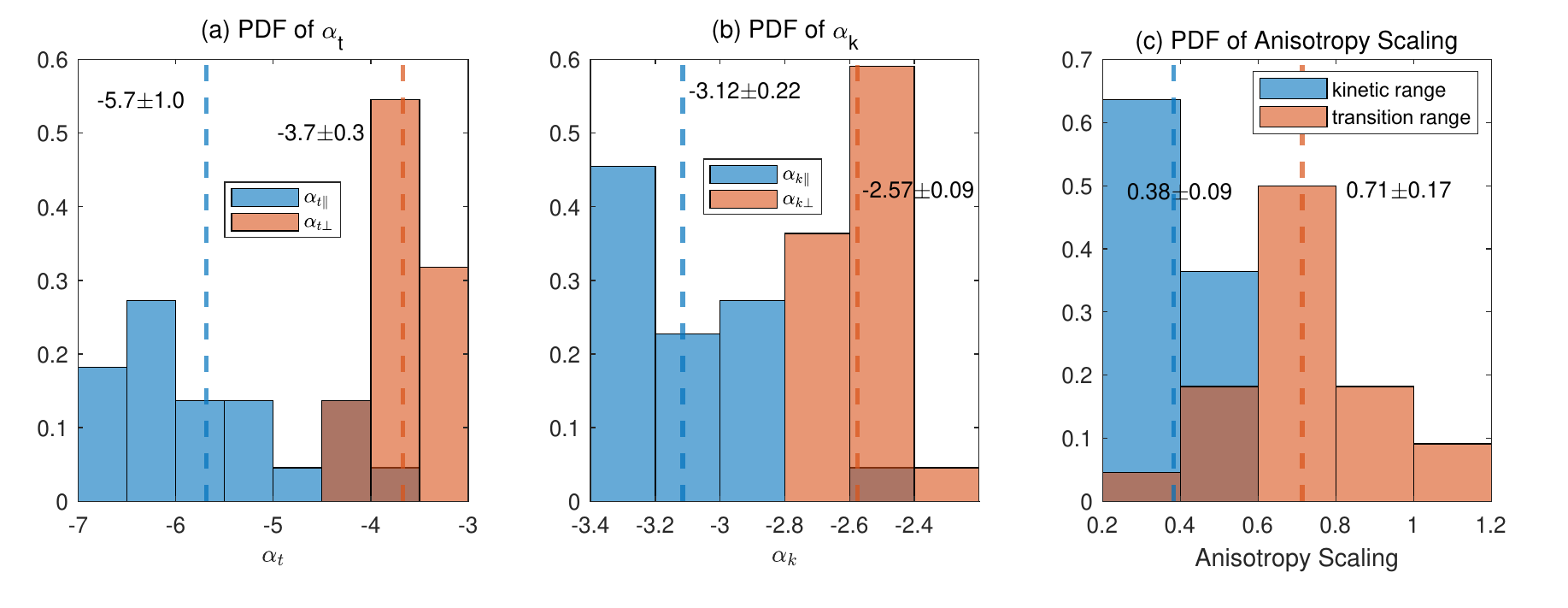}}
    \caption{The statistical result (probability) of the spectral indices in the transition range (a) and kinetic range (b). (c) The statistical result of the wavevector anisotropy. Red indicates the transition range and blue indicates the kinetic range. Dashed lines are the average values.}
    \label{fig4}
\end{figure}

Figure \ref{fig4}(a) and (b) exhibits the statistical results of the spectral anisotropy.\textbf{ Table \ref{tab:1} lists the anisotropy from observation and theoretical predictions.} The parallel direction has the steepest indices, with $\alpha_{t\parallel}=-5.7\pm 1.0$, and $\alpha_{k\parallel}=-3.12\pm 0.22$ for the transition and kinetic ranges. The spectral indices of $\alpha_{t\perp}=-3.7\pm 0.3$ and $\alpha_{k\perp}=-2.57\pm 0.09$ are observed along the perpendicular direction. This result confirms the existence of a transition range signature in both parallel and perpendicular directions, with a trend that the spectra get steeper from the perpendicular direction to the parallel direction. 

Figure \ref{fig4}(c) shows the histograms of the scalings of the wavevector anisotropy. In the transition range along the perpendicular direction, the average scaling is $l_{t\parallel}\sim l_{t\perp}^{0.71\pm0.17}$. The scaling in the kinetic range along the perpendicular direction is $0.38\pm 0.09$, following the relation of $l_{k\parallel}\sim l_{k\perp}^{1/3}$.

\begin{deluxetable*}{cccc}
\tablenum{1}
\tablecaption{Models and Observations of the wavevector anisotropy\label{tab:1}}
\tablewidth{0pt}
\tablehead{
\colhead{Type} & \colhead{Anisotropy Scaling} & \colhead{$P_\perp\ \mathrm{Spectral\ Index}$}& \colhead{Ref.}
}
\startdata
PSP kinetic range& $0.38\pm 0.09$ & $-2.57\pm 0.09$& This Study \\
PSP transition range& $0.71 \pm 0.17$ & $-3.7\pm 0.3$ &This Study\\
Critical-Balanced AW turbulence & 2/3 &-5/3& \citet{Goldreich1995} \\
Critical-Balanced KAW turbulence & 1/3 & -7/3&\citet{Schekochihin2009} \\
Intermittent KAW turbulence & 2/3 &-8/3& \citet{Boldyrev2012} \\
Critical-Balanced ICW turbulence & 5/3 & -11/3&\citet{Schekochihin2019} \\
Tearing-mediated KAW turbulence & $1 \sim 2/3$ & $-3 \sim -8/3$ (depending on the profile of the current sheets)& \citet{Boldyrev2019} \\
\enddata
\end{deluxetable*}

\section{Conclusion and Discussion} \label{sec:conclusion}
In this letter, we present a statistical study of the anisotropy in the kinetic-scale range in the inner heliosphere. By measuring the power spectra along different $\theta_{BV}$, the anisotropy of spectral index and wavevector in the transition range and the kinetic range are investigated. We show that the transition range and the kinetic range have different scalings of anisotropy. The spectral indices varies from $\alpha_{t\parallel}=-5.7\pm 1.3$ to $\alpha_{t\perp}=-3.7\pm 0.3$ in the transition range and $\alpha_{k\parallel}=-2.9\pm 0.2$ to $\alpha_{k\perp}=-2.57\pm 0.07$ in the kinetic range. The wavevector anisotropy exhibits the feature of the KAW turbulence, with the scaling of $k_\parallel \sim k_\perp^{0.7}$ in the transition range and changing to $k_\parallel \sim k_\perp^{1/3}$ in the kinetic range.

\textbf{The observed transition range for the perpendicular spectra is steeper than the cascade models of pure kinetic Alfv\'en waves (-7/3 or -8/3), but could be consistent with dissipation or imbalanced turbulence models. There is a companion paper of Woodham et al. (2021, in prep.) showing that the  magnetic helicity and magnetic compressibility at transition and kinetic range are also consistent with the 1 au observations of KAW turbulence \citep{He2011, Salem2012, kiyani2012enhanced} in \textit{PSP} measurements at 0.17 au. The possible ion-scale dissipation mechanisms are suggested as Landau damping \citep{Howes2008}, cyclotron damping of KAWs \citep{isenberg2019perpendicular} and/or stochastic heating \citep{Chandran2010}, which energize particles in different directions. Although the stochastic heating is stronger when closer to the Sun, there is no direct correlation between the transition range and stochastic heating parameters \citep{Bowen2020inner}. A further study combining particle distribution functions and electric field will help us to investigate the anisotropic dissipation in the inner heliosphere.}

\textbf{Another possible reason for the transition range is the imbalanced turbulence. The dispersive kinetic waves allow nonlinear interaction between the co-propagating wave packets, which can lead to a steepened transition range at ion scales, but the required imbalance for $\alpha_t < -3.5$ is much stronger than the observation \citep{Voitenko2016}. On the other hand, there is a proposed “helicity barrier” from the finite-Larmor-radius MHD in $\beta_p \ll 1$ plasma near the ion scales preventing the energy cascading to the smaller scales \citep{meyrand2020violation}. Only a small portion of energy would leak through the barrier and produce a steep transition range. However, $\beta_p$ and $\beta_e$ are usually larger than 0.1 at 0.17 au, the helicity barrier may not work under this intermediate $\beta$. We can investigate this effect in the future when PSP accesses the lower $\beta$ region of the upper solar atmosphere. }

The ion-scale structures also may contribute the transition range and anisotropy. \citet{Boldyrev2019} predicts that the ion-scale current sheets from the tearing instability could mediate the kinetic Alfv\'enic turbulence to the scalings of $k_\parallel \sim k_\perp^{2/3}$ or  $k_\parallel \sim k_\perp$, but it is difficult to identify such structures with \textit{PSP} observations. \textbf{In addition, a recent observation shows the magnetic fluctuation at sub-ion scale is intermittent with non-Gaussian distribution in the merged dataset \citep{chhiber2021subproton}. How the coherent and intermittent structures contribute to the transition range still need to be clarified.}

\textbf{The perpendicular kinetic range spectral indices measured by PSP are shallower than the -2.8 value measured at 1 au \citep{Sahraoui2009,Chen2010,Alexandrova2012}. At kinetic scale frequencies, the turbulent spectrum at times approaches the level of the noise floor \citep{bowen2020merged}; the frequency dependence of the signal-to-noise ratio may impact estimates of the spectral scaling; however, physical variation in the kinetic range index may be a significant observational signature in constraining kinetic range turbulent dynamics. Our estimate of the anisotropic scaling of the transition range  ($<$ 10 Hz) is largely unaffected by the presence of instrumental noise as the signal-to-noise ratio is sufficiently large. Moreover, towards smaller scales (higher spacecraft-frame frequencies), the turbulence becomes more anisotropic, meaning that $\theta_{BV}$ needs to be measured to greater accuracy to capture the true local parallel spectrum \citep{chen2011anisotropy}. At some point, the limit of the experimental uncertainty on $\theta_{BV}$ is met, resulting in leakage of the perpendicular spectrum into the parallel spectrum. The measurements here suggest that this might be taking place for $f\gtrsim 10$ Hz, however, this does not affect the main results of this Letter since the perpendicular spectrum and wavevector anisotropy measurements do not make use of this high-frequency part of the local parallel spectrum.
 }
 
 Below the transition range, the spectral indices in parallel and perpendicular spectra is similar to the measurements at 1 au \citep{Chen2010} and the -3 parallel spectra is similar to the simulation of \citet{Landi2019}, which proposes a 2D intermittent model at the sub-ion scales. However, the $k_\parallel \sim k_\perp^{1/3}$ anisotropy scaling is inconsistent with the intermittent model. \textbf{There is not yet a complete model to explain all of the features of the spectra measured in this Letter. A unified picture for the anisotropic behaviour in both transition and kinetic ranges remains to be built in the future.}



\appendix 
\section{Piecewise PSD Fitting for Transition Ranges}\label{App}
To determine the frequency range of the transition range, we divide the $\rm{PSD}(f)$ into three sections, which connect one another at the points of $f_1$ and $f_2$, respectively. The inertial range is from 0.1 Hz to $f_1$, the transition range is from $f_1$ to $f_2$, and the kinetic range is from $f_2$ to 90 Hz. We implement linear fitting to each range and get a piecewise linear fitting function in the log-log space:
\begin{equation}
\log_{10}\mathrm{PSD}_{fit}(f;f_1,f_2)=\left\{
\begin{array}{rcl}
\alpha_i\log_{10}f+b_i & & {0.1\ \mathrm{Hz} < f \leq f_1}\\
\alpha_t\log_{10}f+b_t & & {f_1 < f \leq f_2}\\
\alpha_k\log_{10}f+b_k & & {f_2  < f \leq 90\ \mathrm{Hz}}
\end{array} \right.
\end{equation}
Then we compute the deviation function:
\[\mathrm{Dev}(f_1,f_2)=\sum^n_{i=1}[\log_{10}\mathrm{PSD}_{fit}(f_i;f_1,f_2)-\log_{10}\mathrm{PSD}(f_i)]^2,\]
$n$ is the total number of the frequencies. We search the best $f_1$ and $f_2$ to minimize the deviation function and finally we get the frequency range $[f_1, f_2]$ as the transition range.

\acknowledgments
We thank the NASA \textit{Parker Solar Probe} Mission and the FIELDS and SWEAP teams for use of data. D.D. and J.S.H. are supported by NSFC under 41874200 and CNSA under D020301 and D020302. L.D.W. was supported by the STFC consolidated grant ST/S000364/1 to Imperial College London. C.H.K.C. is supported by STFC Ernest Rutherford Fellowship ST/N003748/2 and STFC Consolidated Grant ST/T00018X/1. The FIELDS and the SWEAP experiment on the Parker Solar Probe spacecraft was designed and developed under NASA contract NNN06AA01C. The authors acknowledge the extraordinary contributions of the Parker Solar Probe mission operations and spacecraft engineering teams at the Johns Hopkins University Applied Physics Laboratory. PSP data is available on SPDF (https://cdaweb.sci.gsfc.nasa.gov/index.html/).

%


\bibliography{reference}
\bibliographystyle{aasjournal}



\end{document}